\documentclass[12pt,a4paper]{article}
\usepackage{amsmath}
\usepackage{graphicx}
\usepackage{epstopdf}
\usepackage{color}

\usepackage{cite}

\newcommand{\be}{\begin{equation}}
\newcommand{\en}{\end{equation}}

\renewcommand{\vec}[1]{\boldsymbol{#1}}

\title{Tuning \color{black} the pull-in \color{black} instability of soft dielectric \color{black} elastomers \color{black} through loading protocols}

\author{
Y.\color{black}P. \color{black} Su$^{1,2}$, W.\color{black} Q. \color{black} Chen$^{1}$, M. Destrade$^{2,1}$\\[24pt]
$^1$ Department of Engineering Mechanics,\\
Zhejiang University, Hangzhou 310027, P.R. China\\[12pt]
$^2$ School of Mathematics, Statistics and Applied Mathematics, \\
NUI Galway, University Road, Galway, Ireland 
\\[12pt]
}

\date{}
\begin{document}

\numberwithin{equation}{section}

\maketitle

\begin{abstract}
\color{black} Pull-in (or electro-mechanical) \color{black} instability occurs when a drastic decrease in the thickness of a dielectric elastomer results in electrical breakdown, which limits the applications of dielectric devices. 
Here we derive the \color{black} criterions \color{black} for determining the pull-in instability of dielectrics actuated by different loading methods: voltage-control, charge-control, fixed pre-stress and fixed pre-stretch, \color{black} by analyzing the free energy of the actuated systems\color{black}. 
The \color{black} Hessian criterion \color{black} identifies a maximum in the loading curve beyond which the elastomer will stretch rapidly and lose stability, and can be seen as a path to failure.
We present numerical calculations for neo-Hookean ideal dielectrics, and obtain the maximum allowable actuation stretch of a dielectric before failure by electrical breakdown. 
We find that applying a fixed pre-stress or a fixed pre-stretch to a charge-driven dielectric may decrease the stretchability of the elastomer, a scenario which is the opposite of what happens in the case of a voltage-driven dielectric. 
Results show that a reversible large actuation of a dielectric elastomer, free of the \color{black} pull-in instability\color{black}, can be achieved by tuning the actuation method.

\end{abstract}


\newpage


\section{Introduction}


As some of the most promising soft smart materials capable of performing large actuation deformation in fast response to electrical simulation, dielectric elastomers (DEs) attract considerable attention from academia and industry alike, with potential applications as actuators, sensors, flexible electronic devices, soft robots, energy harvesters, etc.
Their working principle is that in the presence of an electric field, large electrostrictive stresses can be generated and make them  thicker or thinner, and consequently (because they are incompressible), shrink or stretch to large extents \cite{Zhao2008, Dorfmann2006, Zhao2010, Reitz2008, Suo2008}.

\begin{figure}[ht!]
\centering
\includegraphics[width=0.9\textwidth]{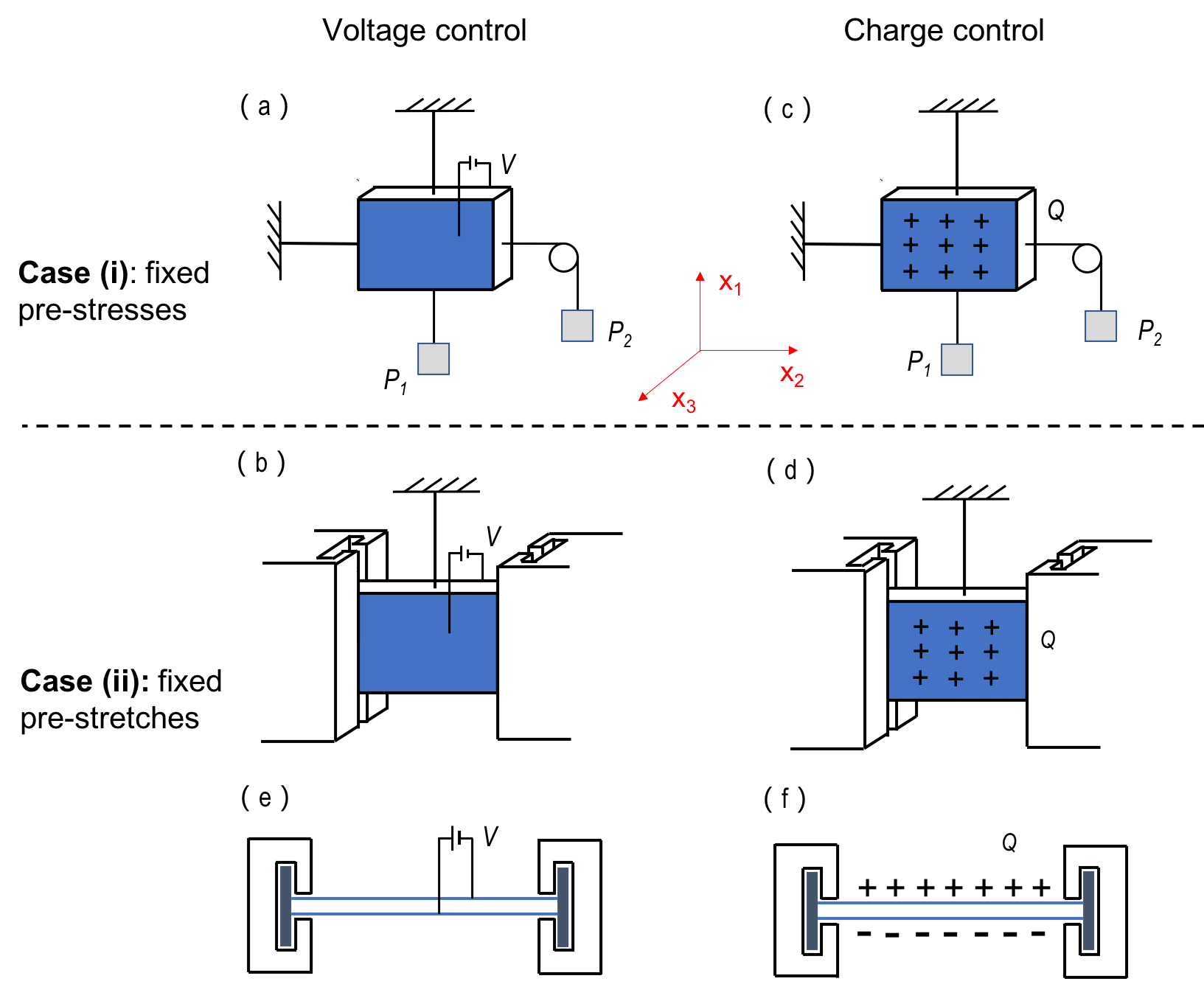}
\caption{
{\small
DEs coated with two compliant electrodes activated by fixed voltage (left column) and fixed charge (right column). 
(a), (c): The plates are pre-stressed by two weights $P_1$ and $P_2$. 
(b), (d): The plates are pre-stretched up to a fixed value in one in-plane direction.
(e), (f): Top view of dielectric plates of (b) and (c), respectively. } 
}
\label{fig1}
\end{figure}

A widely adopted mechanism for creating the electrical activation of a DE is to apply a voltage through  two compliant electrodes glued on the faces of the DE plate; this is the so-called \emph{voltage-controlled method}, see Figure \ref{fig1}a,b and References \cite{Koh2011, Bortot2015, Zurlo2018} for example. 
Generally, the application of a constant voltage causes expansion in area and reduction in thickness of the plate, and as a result leads to an increase in the electric field. 
\color{black}Pull-in instability \color{black} occurs once the applied voltage reaches a threshold value, at which the plate starts expanding rapidly towards a much higher value of stretch. 
Zhao and Suo \cite{Zhao2007} showed that this electro-mechanical instability occurs when the Hessian matrix of the free energy of the whole system ceases to be positive definite, \color{black} so that the equilibrium -- an extremum of the free energy -- is no longer a minimum of the free energy. \color{black}
In practice, the \color{black} pull-in instability \color{black} of the DE may induce electrical breakdown \cite{Stark1955, Zurlo2017} or extensional buckling \cite{Su2018a} of the elastomer, and may thus restrict the actuation stretch. 
Several methods, such as applying pre-stress and tuning material stretchability and dielectricity, have been proposed to harness the \color{black} pull-in \color{black} instability of the DE efficiently \cite{Huang2012,Zhang2017,Su2018b}.

\emph{Charge-controlled actuation} (Figure \ref{fig1}c,d) is another effective method to generate an electric field to activate a deformable capacitor made of soft DE, by spraying charges on the plate with \cite{Lu2014} or without \cite{Keplinger2010} electrodes. 
In contrast to the voltage-controlled actuated DE, it is found that a large reversible actuation can be obtained for the charge-controlled actuated DE without \color{black} pull-in \color{black} instability \cite{Keplinger2010, Li2011}. 
However, the physical mechanism at play to eliminate this instability is still not completely understood.

In this paper, we investigate the nonlinear responses of voltage- and charge-controlled DE plates in turn. 
In particular we formulate the Hessian stability criterion for a charge-controlled DE plate with different boundary conditions, \color{black} by analyzing the variations of the free energy of the system, \color{black} and use it to study the \color{black} pull-in \color{black} instability of the elastomer. 
We show that from an energy perspective, charge-controlled actuation of neo-Hookean ideal dielectric plates is always stable and that the \color{black} pull-in \color{black} instability may be suppressed. 

\color{black}Here we focus on the pull-in instability of a DE plate, a homogeneous deformation mode of instability, which is based on the variations of the free energy and does not account for the plate thickness. It has been shown that pull-in instability may also induce wrinkling instability, which depends on the thickness of the plate \cite{Su2018a}. 
To predict this inhomogeneous instability mode, the incremental theory of electroelasticity \cite{Su2018a, Dorfmann2010, Dorfmann2014, Bortot2018} should be employed, but this is beyond the scope of the current paper.\color{black}

We consider two scenarios: (i) one where the DE plate is subject to a fixed pre-stress, typically by applying weights on its lateral faces; and (ii) another where the plate is held at a fixed pre-stretch in one in-plane direction.
The maximal allowable actuation areal expansion of each case is investigated, by comparing the critical value of voltage or electric displacement for the onset of \color{black} pull-in \color{black} instability with that for electrical breakdown. 
The results indicate that a DE plate may increase its actuation area more in Case (ii) than in Case (i), for voltage- and for charge-controlled actuations.


\color{black}
\section{Energy stability analysis}


Consider a homogeneously deformed incompressible DE plate subject to mechanical loadings $P_1$, $P_2$ along the $x_1$ and $x_2$ directions and a homogeneous nominal electric field $E_0$ along the thickness of the plate, with  associated nominal stress $s_1$, $s_2$ and nominal electric displacement $D_0$, respectively. 
The plate is assumed to be traction-free on its faces, i.e., $s_3=0$.

The free energy (per unit volume) of the deforming system reads \cite{Zhao2007}
\begin{equation}\label{connection0}
G=W\left(\lambda_1,\lambda_2,D_0\right)-s_1\lambda_1-s_2 \lambda_2-E_0D_0,
\end{equation}
where $W\left(\lambda_1,\lambda_2,D_0\right)$ is the energy function of the DE plate, $\lambda_i$ is the principal stretch along the $x_i \ (i=1,2,3)$ direction, with $\lambda_3=\lambda_1^{-1}\lambda_2^{-1}$ due to the incompressibility of the material.

On the other hand, we may define another energy function of the elastomer as $\Omega = \Omega(\lambda_1, \lambda_2, E_0)$, through the partial Legendre transform \cite{Dorfmann2006}
\begin{equation}\label{connection}
W\left(\lambda_1,\lambda_2,D_0\right)=\Omega\left(\lambda_1,\lambda_2,E_0\right)+E_0D_0.
\end{equation}

The first variation of Eq. \eqref{connection} yields \cite{Hannah}
\begin{equation}
W_{\lambda_1}\delta\lambda_1+W_{\lambda_2}\delta\lambda_2+W_{D_0}\delta D_0=
\Omega_{\lambda_1}\delta\lambda_1+\Omega_{\lambda_2}\delta\lambda_2+\Omega_{E_0}\delta E_0+E_0\delta D_0+D_0\delta E_0,
\end{equation}
leading to the identities
\begin{align}\label{connection1}
W_{\lambda_1}=\Omega_{\lambda_1}, \quad W_{\lambda_2}=\Omega_{\lambda_2}, \quad E_0=W_{D_0}, \quad D_0=-\Omega_{E_0}.
\end{align}
Note that here and throughout the paper, subscripts of $W$ and $\Omega$ denote partial derivatives. 

Equilibrium corresponds to the vanishing of the first variation of the free energy of the system, i.e. $\delta G=0$, which using Eq. \eqref{connection1} reads
\begin{align}\label{constitutive}
s_1=W_{\lambda_1}=\Omega_{\lambda_1}, \quad s_2=W_{\lambda_2}=\Omega_{\lambda_2}, \quad E_0=W_{D_0}, \quad D_0=-\Omega_{E_0}.
\end{align}

A thermodynamic analysis of the system \cite{Zhao2007} says that the second variation of its free energy must be positive for the equilibrium to be stable, i.e.,
\begin{multline}\label{secondvariation}
\delta^2G= W_{\lambda_1\lambda_1}\left(\delta\lambda_1\right)^2+2W_{\lambda_1\lambda_2}\delta\lambda_1\delta\lambda_2+
2W_{\lambda_1D_0}\delta\lambda_1\delta D_0  \\ +
W_{\lambda_2\lambda_2}\left(\delta\lambda_2\right)^2+
2W_{\lambda_2D_0}\delta\lambda_2\delta D_0+W_{D_0D_0}\left(\delta D_0\right)^2>0.
\end{multline}
Now the \emph{pull-in instability} of the elastomer occurs when this second variation ceases to be positive and 
\begin{align}
\delta^2G=0,
\end{align}
i.e., when $\vec H_e$, the Hessian matrix of $W$,
\begin{equation}
\vec H_e=
\left[ \begin{matrix}
W_{\lambda_1\lambda_1} & W_{\lambda_1\lambda_2} & W_{\lambda_1 D_0} \\[10pt]
W_{\lambda_1\lambda_2} & W_{\lambda_2\lambda_2} & W_{\lambda_2D_0}  \\[10pt]
W_{\lambda_1D_0}  & W_{\lambda_2D_0}  & W_{D_0D_0}  
\end{matrix}\right],
\end{equation}
ceases to be positive definite \cite{Zhao2007}.

On the other hand, by eliminating the terms involving $\delta^2\lambda_1$ ,$\delta^2\lambda_2$, $\delta^2D_0$, $\delta^2E_0$ and using the connection $\delta D_0=-\left(\Omega_{\lambda_1E_0}\delta\lambda_1+\Omega_{\lambda_2E_0}\delta\lambda_2+\Omega_{E_0E_0}\delta E_0\right)$ found from Eq. \eqref{connection1}$_4$, we find that the second variation of Eq. \eqref{connection} can be written as
\begin{multline}\label{connection2}
W_{\lambda_1\lambda_1}\left(\delta\lambda_1\right)^2+2W_{\lambda_1\lambda_2}\delta\lambda_1\delta\lambda_2 + 2W_{\lambda_1D_0}\delta\lambda_1\delta D_0 \\
+ W_{\lambda_2\lambda_2}\left(\delta\lambda_2\right)^2+
2W_{\lambda_2D_0}\delta\lambda_2\delta D_0+W_{D_0D_0}\left(\delta D_0\right)^2 \\
= \Omega_{\lambda_1\lambda_1}\left(\delta\lambda_1\right)^2+2\Omega_{\lambda_1\lambda_2}\delta\lambda_1\delta\lambda_2+\Omega_{\lambda_2\lambda_2}\left(\delta\lambda_2\right)^2-\Omega_{E_0E_0}\left(\delta E_0\right)^2.
\end{multline}
Here we can see that the left-hand side of Eq. \eqref{connection2} is actually the second variation of $G$, the free energy of the system \eqref{secondvariation}, which must be positive for stability. 
Equivalently, the right-hand side of Eq. \eqref{connection2} must be positive for stability.

For the case of an equi-biaxially deformed DE plate, when $s_1=s_2=s$ and $\lambda_1=\lambda_2=\lambda$, we introduce the following reduced energy functions,
\begin{equation}
w\left(\lambda,D_0\right)=W\left(\lambda,\lambda,D_0\right), \qquad
\omega\left(\lambda,E_0\right)=\Omega\left(\lambda,\lambda,E_0\right).
\end{equation}
Correspondingly, Eqs. \eqref{constitutive} and \eqref{connection2} reduce  to
\begin{align}\label{constitutive-equi}
s=w_{\lambda}/2=\omega_{\lambda}/2, \qquad E_0=w_{D_0}, \qquad D_0=-\omega_{E_0},
\end{align}
and
\begin{align}\label{connection2-equi}
& w_{\lambda\lambda}\left(\delta\lambda\right)^2+2w_{\lambda D_0}\delta\lambda\delta D_0+w_{D_0D_0}\left(\delta D_0\right)^2=
\omega_{\lambda\lambda}(\delta\lambda)^2-\omega_{E_0E_0}\left(\delta E_0\right)^2
,
\end{align}
respectively.

In the paper, we perform numerical calculations with the so-called \emph{neo-Hookean ideal dielectric} model for illustration, defined by
\begin{align}\label{energyfunction}
&W\left(\lambda_1,\lambda_2,D_0\right)=\frac{\mu}{2}\left(\lambda_1^2+\lambda_2^2+\lambda_1^{-2}\lambda_2^{-2}-3\right)+\frac{D_0^2}{2\varepsilon\lambda_1^2\lambda_2^2},\notag\\
&w\left(\lambda,D_0\right)=\frac{\mu}{2}\left(2\lambda^2+\lambda^{-4}-3\right)+\frac{D_0^2}{2\varepsilon\lambda^4},\notag\\
&\Omega\left(\lambda_1,\lambda_2,E_0\right)=\frac{\mu}{2}\left(\lambda_1^2+\lambda_2^2+\lambda_1^{-2}\lambda_2^{-2}-3\right)-\frac{\varepsilon\lambda_1^2\lambda_2^2E_0^2}{2},\notag\\
&\omega\left(\lambda,E_0\right)=\frac{\mu}{2}\left(2\lambda^2+\lambda^{-4}-3\right)-\frac{\varepsilon\lambda^4E_0^2}{2}.
\end{align}
where $\mu$ is the initial shear modulus in the absence of electric field (in Pa) and $\varepsilon$ is the permitivity (in F/m) of the elastomer.
\color{black}
\section{Voltage-controlled actuation of a DE plate}



\subsection{Case (i): Fixed \color{black} equi-biaxial \color{black} pre-stress}


We first consider an incompressible DE plate actuated by a fixed voltage $V$ through the thickness and pre-stressed by two weights $P_1$ and $P_2$ \color{black} along the $x_1$ and $x_2$ directions, \color{black} as depicted in Figure \ref{fig1}(a). \color{black} Here we focus on the case of equi-biaxial deformation, with $P_1=P_2=P$. For this problem we adopt the energy function $\omega\left(\lambda,E_0\right)$ to describe the nonlinear behavior of the elastomer. 

From Eqs. \eqref{constitutive-equi} and \eqref{energyfunction}, we obtain  the equilibrium equations for the neo-Hookean ideal  DE plate as
\begin{align}\label{stress1}
\overline{s}=\lambda-\lambda^{-5}-\overline{E}_0^2\lambda^3, \qquad
\overline E_0=\lambda^{-4}\overline D_0,
\end{align}
where $ \overline s=s/\mu$, $\overline E_0=E_0\sqrt{\varepsilon/\mu}$ and $\overline D_0=D_0/\sqrt{\mu\varepsilon}$ are non-dimensional measures of nominal stress, nominal electric field and nominal electric displacement, respectively. 

Vanishing of the right-hand side of Eq. \eqref{connection2-equi} yields the following criterion for the pull-in instability of the elastomer
\begin{equation}
5+\lambda^{6}-3\overline E_0^2\lambda^8=0,
\end{equation}
which is  the Hessian criterion established by Zhao and Suo \cite{Zhao2007}.\color{black}

Experiments show that the actuation of a DE is limited by its dielectric strength, in the sense that the electric field cannot go beyond $\overline E_B=V_B/(h\sqrt{\mu/\varepsilon})$ (dimensionless), where $V_B$ is the critical applied voltage and $h$ is the deformed thickness of the plate, without failing by electrical breakdown \cite{Stark1955, Pelrine2000}. 
For a deformed incompressible DE, we have the relation
\begin{equation}\label{EB}
\overline E_{0B}=\overline E_B/(\lambda_1\lambda_2),
\end{equation} 
where $\overline E_{0B}=V_B/(H\sqrt{\mu/\varepsilon})$ is the nominal measure of $\overline E_{B}$.

\begin{figure}[ht!]
\centering
\includegraphics[width=0.9\textwidth]{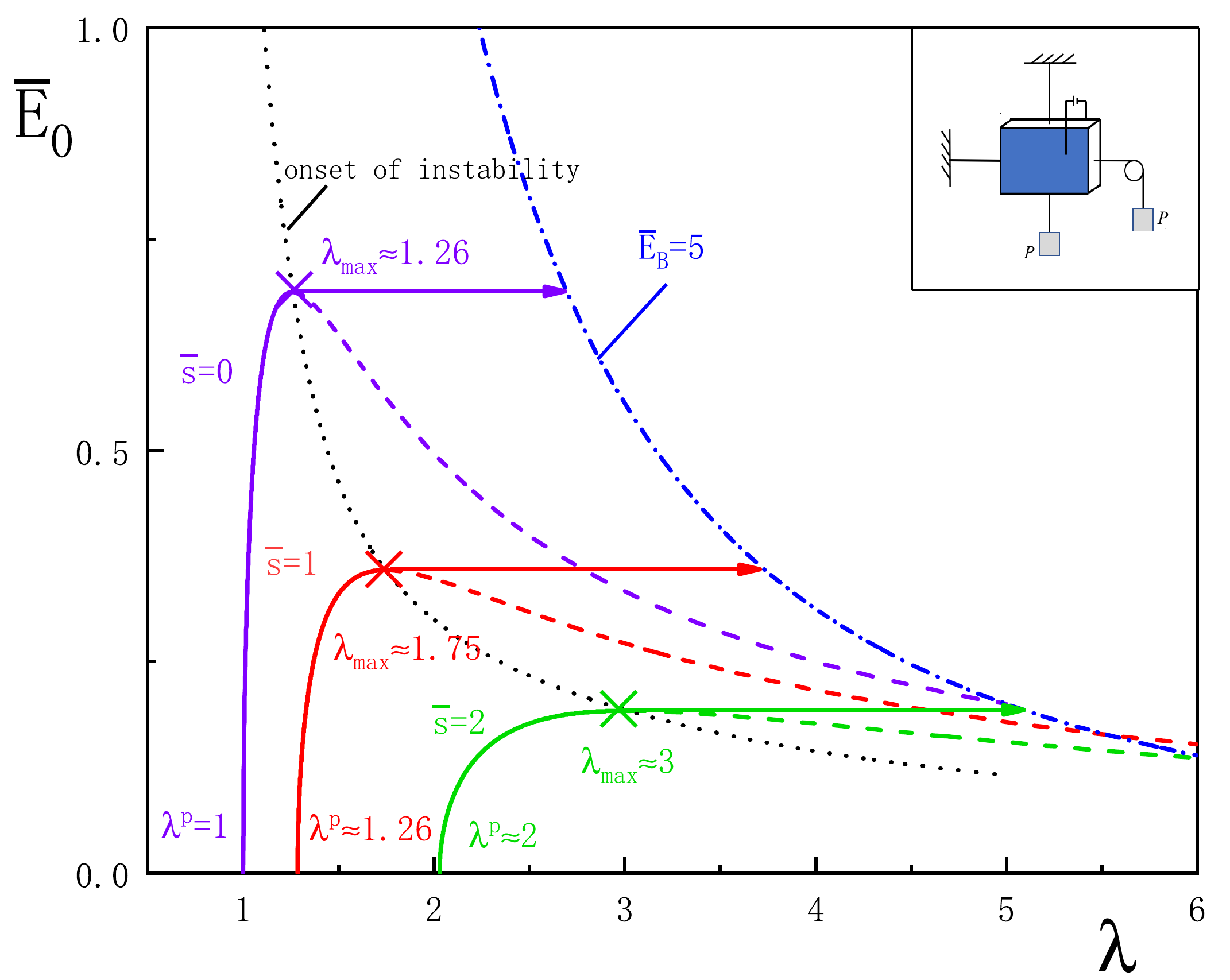}
\caption{
{\small
Effect of fixed pre-stresses ($\color{black}\overline s\color{black}= 0,1,2$) on the behavior of an equi-biaxially deformed DE plate actuated by voltage and pre-stress. 
The critical points for instability are marked by crosses. 
The elastomer may fail by electrical breakdown after \color{black} pull-in \color{black} instability is triggered, as the \color{black} snap \color{black} expansion (arrows) hits the electro-mechanical breakdown (dashed blue curve). The application of pre-stress enhances the stretchability of the elastomer.}
}
\label{fig2}
\end{figure}

Figure \ref{fig2} shows the effect of the equi-biaxial pre-stress on the nonlinear response of the DE plate and its maximal allowable stretch. Solid lines are the $\overline E_0-\color{black}\lambda\color{black}$ curves of the plate with  pre-stresses $\color{black}\overline s\color{black}=0, 1, 2$. Dotted lines corresponds to the onset of \color{black}pull-in \color{black} instability of the elastomer (start of the \color{black}dramatic increase in the area\color{black}). 
Once the instability is triggered, the elastomer  no longer follows  the dashed curve predicted by Eq. \eqref{stress1}, but  experiences a sudden increase (represented by the arrow) in the stretch instead, until fails by electric breakdown represented by the blue dash-dotted curve.

 \color{black} Measurements of strain, electric field, modulus, and dielectric constant of several commonly used dielectric polymers are presented by Pelrine et al. \cite{Pelrine2000} in their Table 1. 
 For this paper, we took the dimensionless breakdown electric field as $\overline E_B=5$, the same value as Koh et al. \cite{Koh2011}.\color{black}

In this study we highlight the so-called \emph{actuation stretch} \cite{Zhao2007}, defined as $\lambda_i^\text{ac}=\lambda_i/\lambda_i^p$ ($i=1,2$), where $\lambda_i^p$ is the pre-stretch in $i$-th direction due to the mechanical loads in the absence of applied voltage. 
Similarly, the actuation expansion in area is defined as $A^\text{ac}=A/A^p$, where $A=\lambda_1\lambda_2$ is the total expansion of area of the deformed plate and $A^p=\lambda_1^p\lambda_2^p$ is the areal expansion due to the mechanical loads. 
Clearly these quantities are linked as $A^\text{ac}=\lambda_1^\text{ac}\lambda_2^\text{ac}$. 

We can see from Figure \ref{fig2} that the maximal allowable actuation stretch of the plate with no mechanical pre-stress ($\color{black}\overline s\color{black}=0$),  is $\color{black}\lambda_\text{max}^\text{ac}\color{black} \simeq 1.26$, with corresponding actuation expansion in  area $\simeq$ 160\%. 
We find that applying a pre-stress enhances the actuation of the DE plate: the maximal allowable actuation stretch of the elastomer is $\lambda_\text{max}^\text{ac} \simeq$ 1.39, 1.5, with actuation expansion in  area  $A_\text{max}^\text{ac} \simeq$ 193 \%, 225\%, respectively, as $\overline s=1,2$ is applied.


\subsection{Case (II): Fixed pre-stretch}


Next we consider an incompressible DE plate with one of the in-plane stretches, $\lambda_2$ say, fixed to a certain amount and subject to a fixed voltage $V$, see Figures \ref{fig1}(b),(e). 
\color{black}The equilibrium equations can then be written as
\begin{equation}\label{constitute2}
s_1=\dfrac{\partial \widetilde \Omega}{\partial {\lambda_1}}, \qquad
D_0=-\dfrac{\partial \widetilde \Omega}{\partial E_0},
\end{equation}
where $\widetilde \Omega(\lambda_1, E_0) = \Omega(\lambda_1, \lambda_2, E_0)|_{\lambda_2 \text{ fixed}}$, and $s_1$ is the in-plane nominal stress in $x_1-$direction ($s_1=0$ here). 
Again the plate is assumed to be traction-free on its main faces ($s_3=0$).

For neo-Hookean ideal dielectrics, Eq. \eqref{constitute2} reads
\begin{align}\label{stress2}
 0 = \lambda_1-\lambda_1^{-3}\lambda_2^{-2}-\overline{E}_0^2\lambda_1\lambda_2^2, \qquad \overline E_0=\lambda_1^{-2}\lambda_2^{-2}\overline D_0,
\end{align}
and pull-in instability happens when the right-hand side of Eq. \eqref{connection2} vanishes, i.e. when
\begin{equation}
3+\lambda_1^4\lambda_2^2(1-\overline E_0^2\lambda_2^2)=0.
\end{equation}
\color{black}

\begin{figure}[ht!]
\centering
\includegraphics[width=0.9\textwidth]{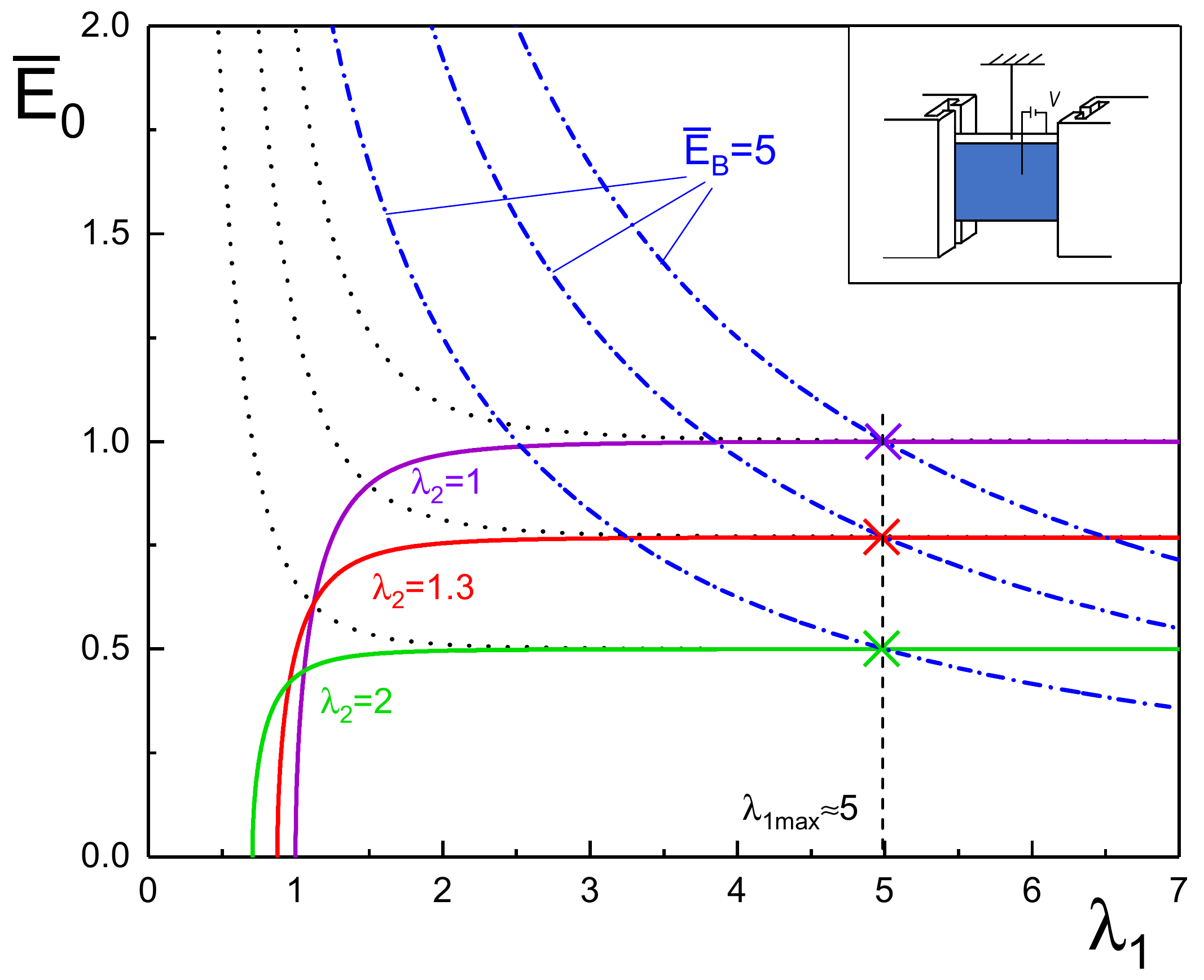}
\caption{
{\small
Effect of a fixed pre-stretch ($\lambda_2 = 1.0, 1.3, 2.0$) on the behavior of a DE plate actuated by voltage. 
The black dotted lines correspond to the onset of \color{black} pull-in \color{black} instability of the elastomer;
they only meet the loading curves asymptotically. 
The critical points for electrical breakdown are marked as crosses. In this case, a large reversible actuation can be obtained without \color{black} pull-in \color{black} instability.}
}
\label{fig3}
\end{figure}

Figure \ref{fig2} displays the effect of the fixed pre-stretch $\lambda_2$ on the nonlinear $\overline E_0 - \lambda_1$ response for a DE plate actuated by voltage. 
Here $\lambda_2$ is held at $1, 1.3, 2$. 
As the voltage applied increases, $\lambda_1$  increases monotonically along the solid lines obtained from Eq. \eqref{stress2}. 
We see that for a pre-stretched DE plate, \color{black} the stretch $\lambda_1$ increases monotonically with the nominal electric field $\overline E_0$,  and that \color{black} there is no intersection between the $\overline E_0-\lambda_1$ curve and the curves of \color{black} pull-in \color{black} instability (except asymptotically as $\lambda_1 \to \infty$).
It follows that a reversible stretch of the elastomer can be obtained without \color{black} pull-in \color{black} instability, as long as it has not reached the electrical breakdown condition.  

Also, the stretchability of a pre-stretched elastomer is much better than in the previous case of equi-biaxial loading (Figure \ref{fig2}), with maximal allowable actual stretch $\lambda_\text{1max} \simeq  5$, which is almost independent of the applied pre-stretch. 
To show this we note that $\lambda_\text{1max}$ is the real root of the equation $\lambda_1^2=\overline E_B^2+\lambda_1^{-2}\lambda_2^{-2}$, see Eqs. \eqref{EB} and \eqref{stress2}$_1$. 
Here $\lambda_1^{-2}\lambda_2^{-2}$ is the square of the stretch through the thickness, so that $\lambda_\text{1max}^{-2}\lambda_\text{2max}^{-2}<1\ll\overline E_B^2$ when electrical breakdown happens. 
As a result, $\lambda_\text{1max} \simeq  \overline E_B$, always. 
As the fixed stretch $\lambda_2$ increases from 1 to 1.3 to 2, $\lambda_\text{1max}^\text{ac}$ and $A_\text{max}^\text{ac}$ both increase from \color{black} 500\% to 568\% to 700\%.\color{black}


\section{Charge-controlled actuation of a DE plate}



\subsection{Case (i): Fixed \color{black} equi-biaxial \color{black} pre-stress}


\color{black}Here we consider a deformed DE plate subject to fixed charge and mechanical loading $P_1$ and $P_2$ as depicted in Figure \ref{fig1}c. For the case of equi-biaxial deformation $P_1=P_2=P$ we use the energy function $w(\lambda,D_0)$ to capture the nonlinear response of the elastomer. 

According to Eqs. \eqref{constitutive-equi} and \eqref{energyfunction}, the governing equations of the neo-Hookean ideal DE plate read
\begin{align}
\overline{s}=\lambda-\lambda^{-5}-\overline{E}_0^2\lambda^3, \qquad
\overline E_0=\lambda^{-4}\overline D_0.
\end{align}

The criterion for instability, seen as the vanishing of the left-hand side of Eq. \eqref{connection2-equi}, reads
\begin{equation}
\lambda^{6}+5(1+\overline E_0^2\lambda^8)=0,
\end{equation}
which has no real root. Thus pull-in (Hessian) instability is suppressed here.\color{black}

\begin{figure}[ht!]
\centering
\includegraphics[width=0.9\textwidth]{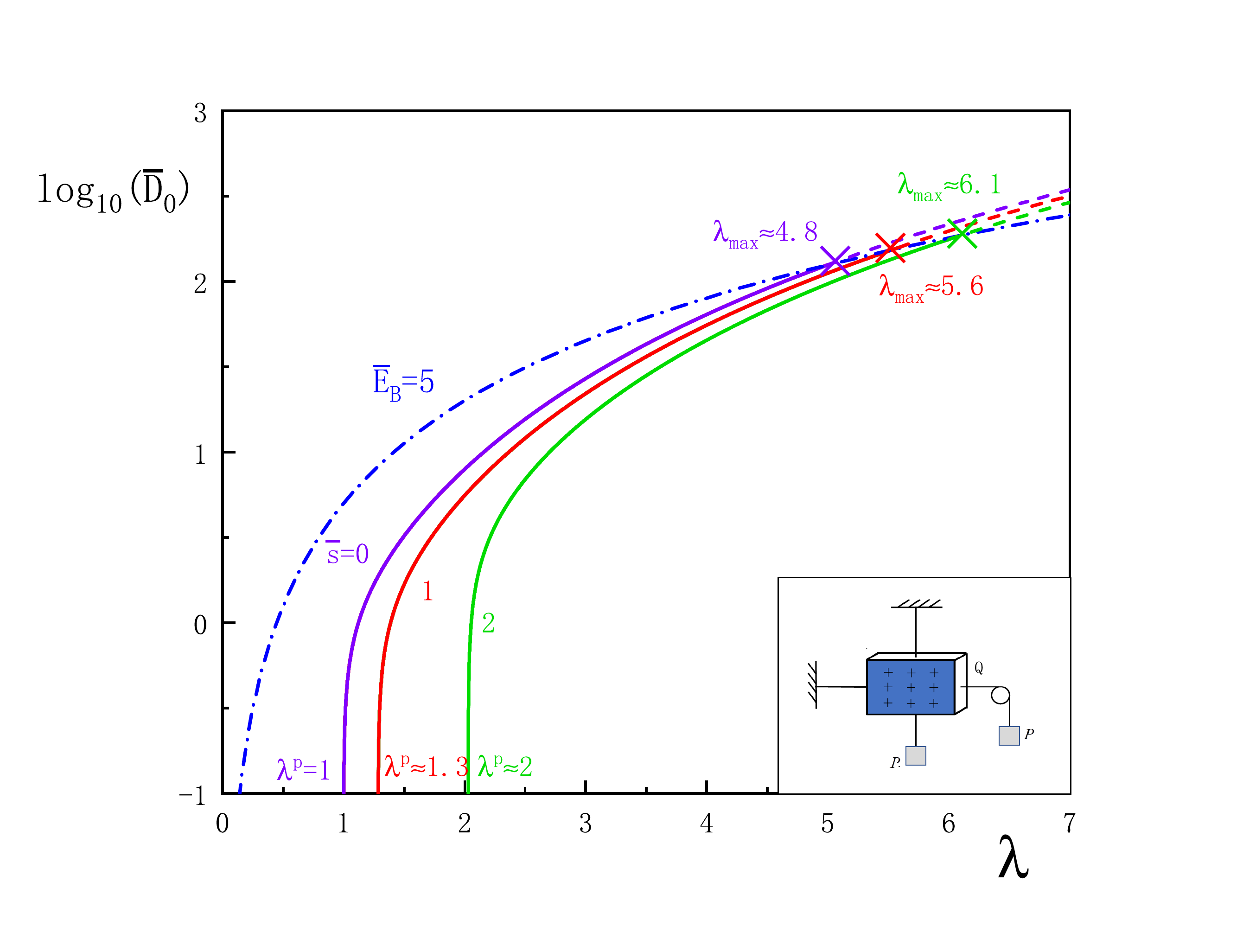}
\caption{
{\small
Effect of fixed pre-stress ($\color{black}\overline s\color{black}=0, 1, 2$) on the behavior of an equi-biaxially deformed DE plate actuated by charges deposited on its faces. 
The critical points for electrical breakdown are marked as crosses. A large reversible actuation can be obtained without pull-in instability, but the application of a fixed pre-stress decreases the actuation stretchability of the elastomer.}
}
\label{fig4}
\end{figure}

Figure \ref{fig4} depicts the effect of a fixed pre-stress ($\overline s=0, 1, 2$) on the nonlinear response of an equi-biaxially deformed DE plate and its maximal allowable stretch. 
Solid lines are the log$_{10}(\overline D_0)-\lambda$ curves. 
As the applied charge increases, the elastomer expands monotonically by a large amount without developing \color{black} pull-in \color{black} instability, until it fails by electric breakdown. 
As the pre-stress $s$ increases from 0 to 1 to 2, $\lambda_\text{max}^\text{ac}$ and $A_\text{max}^\text{ac}$ decrease from 4.8 to 4.3 to 3.05, and from 2,304$\%$ to 1,849$\%$ to 930$\%$, respectively.


\subsection{Case (II): Fixed pre-stretch}


\color{black}Here we adopt the energy function $\widetilde W(\lambda_1, D_0) = W(\lambda_1, \lambda_2, D_0)|_{\lambda_2 \text{ fixed}}$ to write the equilibrium equation of the plate as 
\begin{equation}\label{constitute4}
s_1=\dfrac{\partial \widetilde W}{\partial {\lambda_1}}, \qquad
E_0=\dfrac{\partial \widetilde W}{\partial D_0},
\end{equation}
which for neo-Hookean ideal dielectrics reads
\begin{align}\label{stress4}
0=\lambda_1-\lambda_1^{-3}\lambda_2^{-2}(\overline{D}_0^2+1), \qquad 
 \overline E_0=\lambda_1^{-2}\lambda_2^{-2}\overline D_0.
\end{align}

Vanishing of the left-hand side of Eq.\eqref{connection2} yields the pull-in criterion of this problem as
\begin{equation}
3+3 \overline D_0^2+\lambda_1^4\lambda_2^2=0,
\end{equation}
which can never be achieved.
Thus pull-in (Hessian) instability of the  dielectric elastomer is suppressed for this problem as well.\color{black}

\begin{figure}[ht!]
\centering
\includegraphics[width=0.9\textwidth]{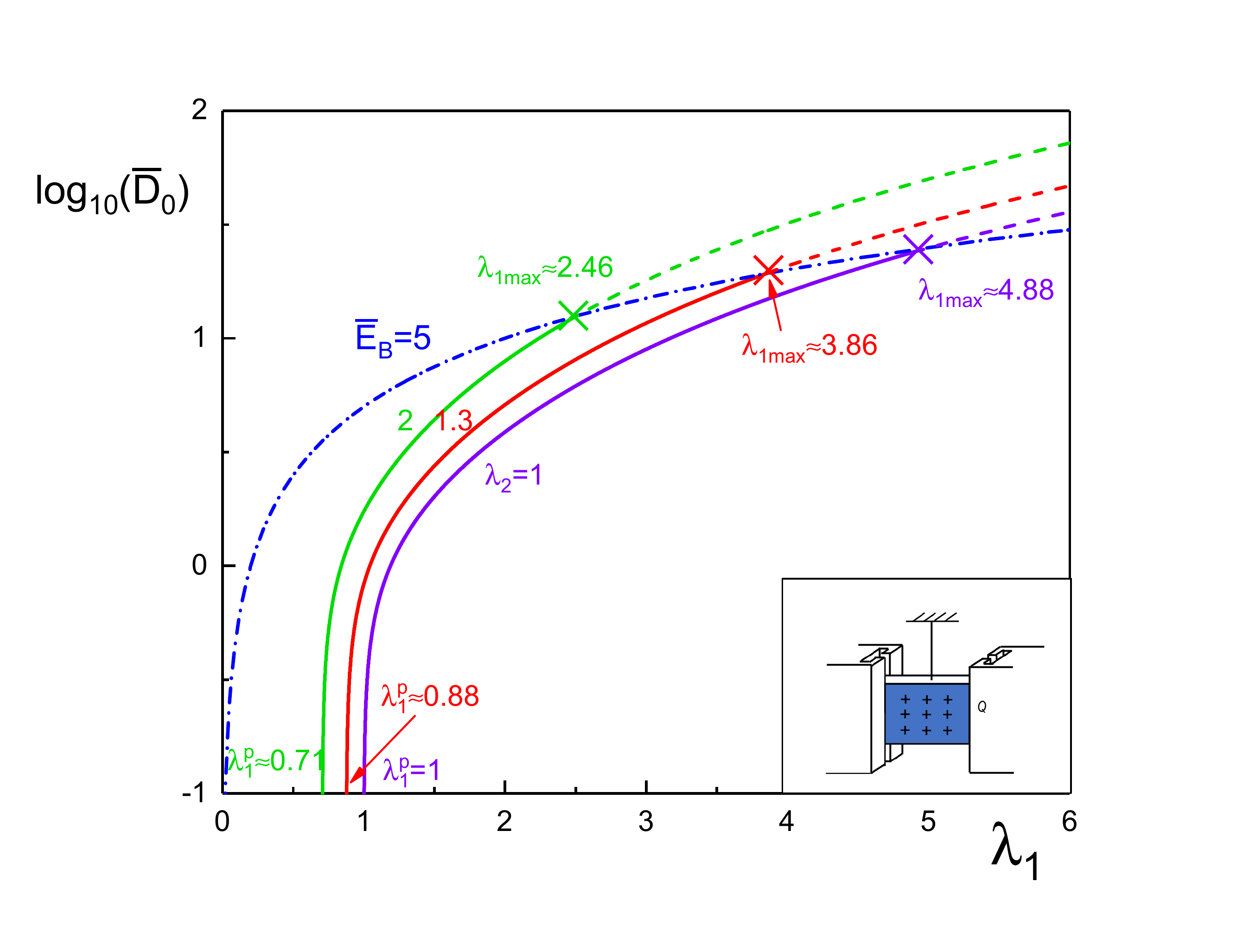}
\caption{
{\small
Effect of a fixed pre-stretch  ($\lambda_2 = 1.0, 1.3, 2.0$) on the behavior of a DE plate actuated by spraying charges on its faces. 
The critical points for electrical breakdown are marked as crosses. A large reversible actuation can be obtained without \color{black} pull-in \color{black} instability, but the application of a fixed pre-stretch may reduce the stretchability of the elastomer.}
}
\label{fig5}
\end{figure}

Figure \ref{fig5} displays the effect of a fixed pre-stretch on the nonlinear response of the DE plate actuated by charge and its maximal allowable stretch. 
The elastomer is initially pre-stretched in $x_2$ direction to a fixed amount ($\lambda_2=1, 1.3, 2$, respectively). 
We can see that for this problem, the \color{black} pull-in \color{black} instability can be suppressed.
On the other hand, the application of a fixed pre-stretch decreases the maximal allowable actuation stretch and maximal actuation areal expansion, which is contrast from the corresponding case of a voltage-controlled actuated DE, displayed in Figure \ref{fig3}. 


\section{Conclusions}

\color{black}
A DE plate contracts in thickness and expands in area when subject to an electric field. Pull-in instability occurs as the applied electric field reaches a critical value, driving the elastomer to thin down rapidly.
On one hand, pull-in instability may lead to electric breakdown. On the other hand,
the large deformation induced by pull-in instability can be exploited in the design of high-performance actuators and sensors because the elastomer may enlarge its area severalfold. The pull-in instability can be suppressed or delayed by pre-stretching the elastomer or enhancing its stiffening property. Dielectric with designed stiffening property may survive the pull-in instability, and large deformation can be obtained without electric breakdown \cite{Zhao2014}. \color{black}

\color{black}In this paper, we explored ways to achieve large deformation of dielectrics by tuning the loading protocol. \color{black} Using the theory of nonlinear electro-elasticity, we investigated the effect of actuation methods on the nonlinear response and \color{black} pull-in \color{black} instability of a DE plate. 
\color{black} We introduced two equivalent energy functions $W(\lambda_1,\lambda_2,D_0)$ and $\Omega(\lambda_1,\lambda_2,E_0)$ to capture the nonlinear response of the DE plate. In fact, due to the connection \eqref{connection0}, the two functions lead to  identical results in the  investigation of instability.\color{black}

We derived the forms of the criterion \color{black}  predicting pull-in instability \color{black} when the elastomer is actuated by voltage or by charge. \color{black} It proved convenient to use $\Omega(\lambda_1,\lambda_2,E_0)$ to study pull-in stability of a voltage-controlled actuated DE plate, and to use $W(\lambda_1,\lambda_2,D_0)$ for the case of a charge-controlled actuated DE. \color{black}
For a voltage-driven DE free to expand equi-biaxially, \color{black} pull-in \color{black} instability may be triggered once a sufficiently large voltage is applied, resulting in electrical breakdown of the elastomer, while the application of a fixed pre-stress can increase its maximal allowable actuation stretch. 
For a voltage-driven DE with constant pre-stretch or a charge-driven DE, a reversible large actuation can be obtained without \color{black} pull-in \color{black} instability. 
The results showed that pre-stressing or pre-stretching a charge-driven DE can decrease the stretchability of the elastomer, which is the opposite of what happens in the case of a voltage-driven DE. 
This study provides a new route for the design of DE actuators with large deformations.

\color{black}Note that we did not consider viscosity \cite{Plante2006, Huang2012} in any way, although experimental observations have exposed its effect on electro-mechanical instability of DE.\color{black}


\section*{Acknowledgments}


This work was supported by a Government of Ireland Postdoctoral  Fellowship from the Irish Research Council (No. GOIPD/2017/1208) and by the National Natural Science Foundation of China (No. 11621062).
WQC and YPS also acknowledge the support from the Shenzhen Scientific and Technological Fund for R$\&$D (No. JCYJ20170816172316775).
MD thanks Zhejiang University for funding research visits  to Hangzhou, and Alain Goriely and Giuseppe Saccomandi for the invitation and support to attend the conference on ``Mathematics \& Mechanics: Natural Philosophy in the 21st Century" at Oxford University in June 2018.



\end{document}